\documentclass[twocolumn,twocolappendix]{aastex631}

\usepackage{graphicx}
\usepackage{xcolor}
\usepackage{placeins}
\usepackage{textgreek}
\usepackage{amsmath}
\usepackage{cleveref}

\begin{document}

\title{Laboratory study of magnetic reconnection in lunar-relevant mini-magnetospheres}

\correspondingauthor{Lucas rovige}
\email{lrovige@ucla.edu}

\author[0000-0002-5267-5574]{Lucas Rovige}
\affiliation{University of California--Los Angeles \\
Los Angeles, CA 90095, USA}

\author{Filipe D. Cruz}
\affiliation{GoLP/Instituto de Plasmas e Fusão Nuclear, Instituto Superior Técnico, Universidade de Lisboa,  \\
1049-001 Lisboa, Portugal}

\author{Robert S. Dorst}
\affiliation{University of California--Los Angeles \\
Los Angeles, CA 90095, USA}

\author{Jessica J. Pilgram}
\affiliation{University of California--Los Angeles \\
Los Angeles, CA 90095, USA}

\author[0000-0002-8439-5518]{Carmen G. Constantin}
\affiliation{University of California--Los Angeles \\
Los Angeles, CA 90095, USA}

\author{Stephen Vincena}
\affiliation{University of California--Los Angeles \\
Los Angeles, CA 90095, USA}

\author{Fábio Cruz }
\affiliation{GoLP/Instituto de Plasmas e Fusão Nuclear, Instituto Superior Técnico, Universidade de Lisboa,  \\
1049-001 Lisboa, Portugal}

\author{Luis O. Silva}
\affiliation{GoLP/Instituto de Plasmas e Fusão Nuclear, Instituto Superior Técnico, Universidade de Lisboa,  \\
1049-001 Lisboa, Portugal}

\author[0000-0001-5489-9144]{Christoph Niemann}
\affiliation{University of California--Los Angeles \\
Los Angeles, CA 90095, USA}

\author[0000-0003-1675-5910]{Derek B. Schaeffer}
\affiliation{University of California--Los Angeles \\
Los Angeles, CA 90095, USA}



\begin{abstract}

Mini-magnetospheres are small ion-scale structures that are well-suited to studying kinetic-scale physics of collisionless space plasmas. Such ion-scale magnetospheres can be found on local regions of the Moon, associated with the lunar crustal magnetic field. In this paper, we report on the laboratory experimental study of magnetic reconnection in laser-driven, lunar-like ion-scale magnetospheres on the Large Plasma Device (LAPD) at the University of California - Los Angeles. In the experiment, a high-repetition rate (1 Hz), nanosecond laser is used to drive a fast moving, collisionless plasma that expands into the field generated by a pulsed magnetic dipole embedded into a background plasma and magnetic field. The high-repetition rate enables the acquisition of time-resolved volumetric data of the magnetic and electric fields to characterize magnetic reconnection and calculate the reconnection rate. We notably observe the formation of Hall fields associated with reconnection. Particle-in-cell simulations reproducing the experimental results were performed to study the micro-physics of the interaction. By analyzing the generalized Ohm's law terms, we find that the electron-only reconnection is driven by kinetic effects, through the electron pressure anisotropy. These results are compared to recent satellite measurements that found evidence of magnetic reconnection near the lunar surface. 

\end{abstract}


\section{Introduction} \label{sec:intro}
\begin{figure*}[ht!]
\plotone{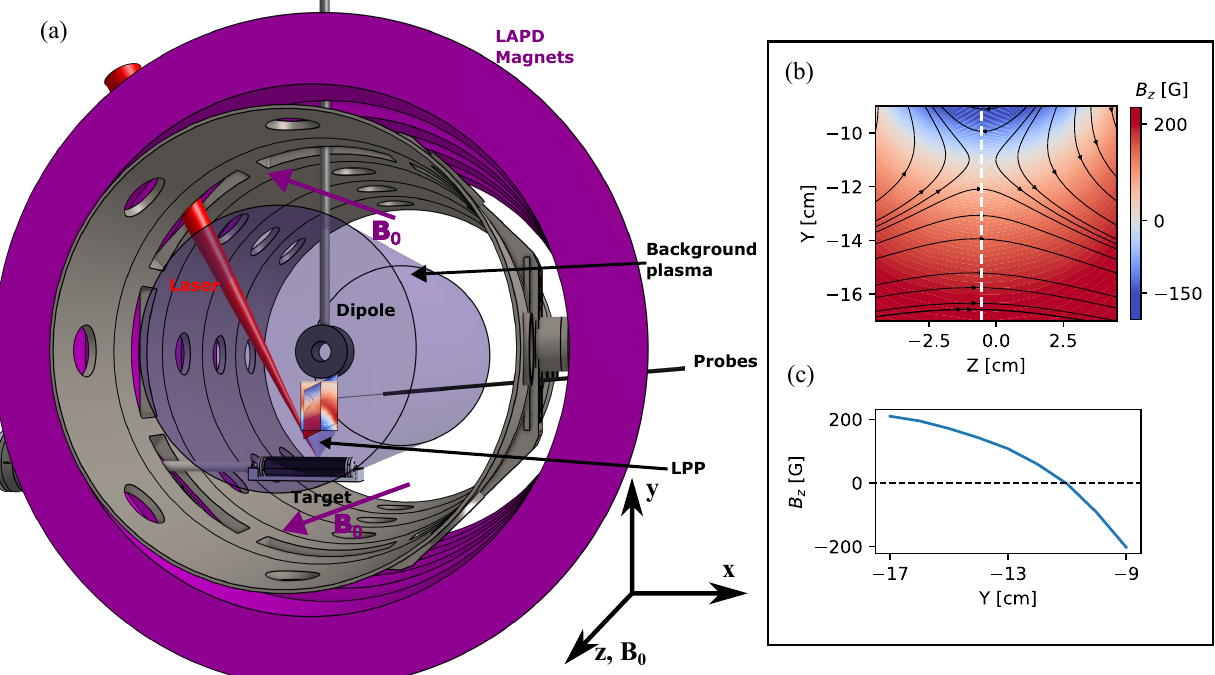}
\caption{ (a) Schematic of the experimental setup.  The center of the dipole electromagnet is located at $\{x,y,z\}=\{0,0,0\}$ cm. (b) Ambient magnetic field $B_z$ from the pulsed dipole and constant LAPD field along the z-axis, measured in the YZ-plane (red-blue) with magnetic field lines (black). (c) Lineout of $B_z$ along the dashed white line in (b). 
\label{fig:setup}}
\end{figure*}

A mini-magnetosphere is a magnetosphere that forms on a length-scale of the order of the ion inertial length. Because of their small system scale, they constitute an ideal means of studying kinetic-scale physics problems. Such ion-scale magnetospheres can be observed in low altitude (10s-100s km) regions of the Moon \citep{lin98,futana06,halekas08,wieser10,lue11,bamford12} and on Mars \citep{fan17}, associated with local crustal magnetic anomalies, while induced mini-magnetospheres can form around initially unmagnetized bodies such as comets \citep{nilsson15}. They also have been proposed for protection of spacecrafts from ionizing radiation \citep{bamford14} and spacecraft propulsion \citep{moritaka12}. The efforts to understand the interaction of the solar wind with these kinetic-scale objects have also been supported by numerical work \citep{omidi02,gargate08,bamford08,bamford16,cruz17} as well as laboratory experiments \citep{bamford12,shaikhislamov13,shaikhislamov14,schaeffer22}. \par
In the past decade, the fundamental role of kinetic effects and electron dynamics in magnetic reconnection has attracted much attention associated with 
\textit{in-situ} observations of electron-only reconnection events by the Magnetospheric Multiscale Mission (MMS) mission \citep{burch16,wilder17,phan2018}, numerical simulations \citep{shay98,chacon07,jain12,hesse14,sharma19} and laboratory experiments \citep{kuramitsu18,yamada18,sakai2022}. In this context, mini-magnetospheres provide a compact environment with typically low ion magnetization, ideal for studying electron-scale magnetic reconnection. Indeed, recent work from \citet{sawyer23} showed the first observational evidence of magnetic reconnection in a lunar mini-magnetosphere using data from the THEMIS-ARTEMIS mission \citep{angelopoulos2014}. These results also showed a clear demagnetization of the ions, thus indicating a probable electron-only reconnection process. However, satellite data at low lunar altitudes necessary to access these mini-magnetospheres are scarce and limited. \par
Laboratory experiments can thus be an invaluable tool to fill in this gap and help understand the physics of magnetic reconnection in lunar and other mini-magnetospheres. We have developed an experimental platform to study ion-scale magnetospheres on the Large Plasma Device (LAPD) at UCLA. This platform operates at a relatively high repetition-rate (1/3 Hz) and combines an ambient magnetized plasma provided by the LAPD, a pulsed magnetic dipole, and a fast, laser-driven plasma flow generated by an energetic laser impinging on a solid target. In a previous proof-of-principle experiment on the platform and related numerical simulations, we demonstrated the formation of a mini-magnetosphere through the observation of a magnetopause and associated current structures \citep{schaeffer22,cruz22}. \par

In this paper we report on the experimental observation of magnetic reconnection in ion-scale magnetospheres on the LAPD, using a field configuration where the dipole and the LAPD fields are anti-parallel on axis. The electric and magnetic fields are measured in a 3D volume using motorized probes, and we observe Hall fields as well as evidence of electron-only reconnection. The experiment is compared to particle-in-cell (PIC) simulations that allow us to study the kinetic physics driving reconnection in mini-magnetospheres through the evaluation of all the terms from the generalized Ohm's law. 
By comparing important dimensional parameters, we show that these results are relevant to similar observations of reconnection in lunar mini-magnetospheres from \citet{sawyer23}.

The paper is set up as follows. Section \ref{sec:setup} presents the experimental set-up. Experimental results are presented in Sec.~\ref{sec:res} and simulation results are shown in Sec.~\ref{sec:simulations}. The evaluation of the Ohm's law terms and the comparison to lunar mini-magnetospheres are discussed in Sec.~\ref{sec:disc}.

\section{Experimental setup} \label{sec:setup}

\begin{figure*}[ht!]
\plotone{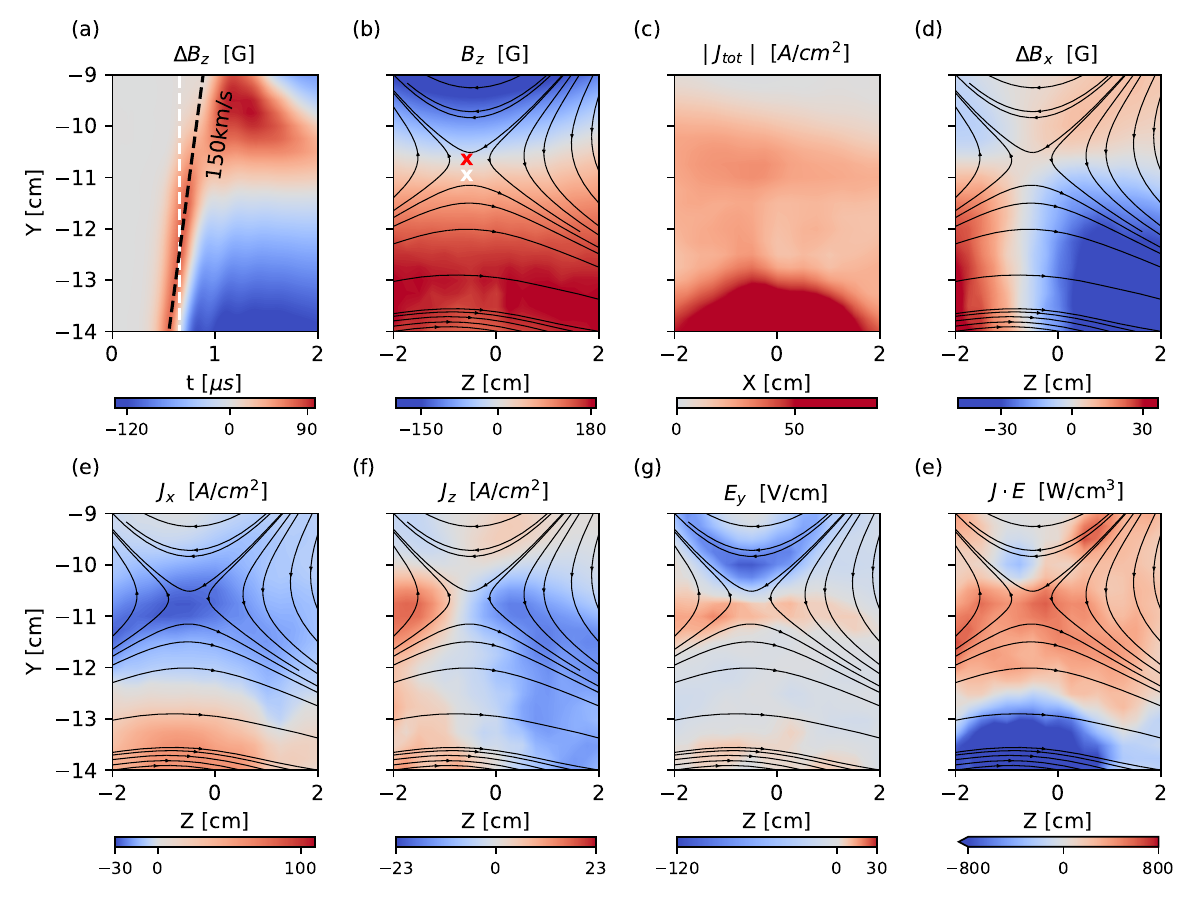}
\caption{ (a) Streak plot along the y-axis of the dynamic magnetic field in the z direction $\Delta B_z = B_{z} - (B_0 + B_{z,dip})$ at $\{x,z\} = \{-0.5,-0.5\}$~cm corresponding to the center of the LPP.  All the data planes shown in the subsequent plots are taken at one of these positions and at a time $t = 0.65$~$\mu$s marked by the white dashed line. The black dashed line highlights the driver propagation speed. (b) Total magnetic field $B_z$ and magnetic field lines (black arrows) in the Y-Z plane. The white cross indicates the initial position of the magnetic null point, and the red cross indicates its position at $t = 0.65$~$\mu$s. (c) Amplitude of the total current density in the Y-X plane, retrieved from Ampere's Law. (d) Dynamic out-of-plane magnetic field $\Delta B_x$ in the Y-Z plane. (e) Out-of-plane current density $J_x$ in the Y-Z plane. (f) Current density $J_z$ in the z direction in the Y-Z plane. (g) Electrostatic field $E_y = - (\nabla \Phi_p)_y$ in the y direction in the Y-Z plane. (h) Dissipation of energy of the electromagnetic field into the plasma $\mathbf{J}\cdot\mathbf{E}$ in the Y-Z plane.
\label{fig:fig2}}
\end{figure*}

The experiment was carried out on the Large Plasma Device (LAPD) at UCLA operated by the Basic Plasma Science Facility (BaPSF) and combined a large-scale (50 cm diameter) ambient magnetized plasma, a pulsed dipolar magnetic field generated by an intense current, and a fast, laser-driven plasma flow. The LAPD \citep{lapd16} is a linear-pulsed discharge device consisting of a 20-m long, 1-m diameter vacuum vessel that can generate a long-lived ($\sim$15~ms) plasma at up to 1 Hz repetition rate. The machine generates a longitudinal (along z) magnetic field that can be varied from 200~G to 2500~G through large inductive magnets that surround it, and variable plasma densities in the range ($10^{11}-10^{13}$ cm$^{-3}$) with typical electron temperatures of $T_e \sim 5-10$ eV and ion temperatures of $T_i \sim 1$ eV. 

Figure~\ref{fig:setup}a shows a schematic of the experimental setup. A 14~cm outer diameter and 4~cm inner diameter pulsed electromagnet is placed at the center of the LAPD ($\{x,y,z\}=\{0,0,0\}$) and generates a dipolar magnetic field with a magnetic moment that can reach up to 550~A/m$^2$ with a 3.8~kA driving current.  The magnet is water cooled to allow for repetition rates up to 1~Hz. A motorized cylindrical graphite target, inserted along the x-axis, is centered below the dipole electromagnet at $y=-30$ cm.\par

The driver beam (12~J over 20~ns) is sent to the chamber from the top at a 30$^\circ$ angle from the target surface normal and focused to an intensity of $I_0 \sim 5\times10^{11}$ W/cm$^2$ onto the graphite target. This generates a fast plasma flow along the y-axis that expands into the background plasma at a speed of 150~km/s, transverse to the background field. The laser, as well as the LAPD and the pulsed magnet, operate at a 1/3~Hz repetition rate, and the target is translated and rotated between every shot so that the laser always hits a fresh surface. The laser was timed to fire at the peak of the dipole field, and the experiment lasted for a few $\mu$s which is much shorter than the lifetime of the ambient plasma ($\sim$10~ms) as well as the characteristic time evolution of the pulsed dipole ($\sim$0.5~ms).  \par 
In the experiment, using hydrogen as ambient gas fill, the background plasma density is set to $n_e = 1\times10^{13}$ cm$^{-3}$, yielding an ion inertial length $d_i = 7.2$ cm. The interaction between the laser-produced plasma (LPP) consisting primarily of $C^{+4}$ ions \citep{schaeffer16} and the ambient $H^+$ plasma is largely collisionless. The ion collisional mean-free-path is much larger than the system size ($\lambda_{ii} > 1$ m), and both the electron-ion and electron-electron collision time are larger than the electron gyroperiod ($\omega_{ce}\tau_{ie} > 23$ and $\omega_{ce}\tau_{ee} > 10$, respectively).   The LAPD field is set to $B_0$ = 300~G. By changing the current flow direction in the pulsed electromagnet, the dipole field is set to be anti-parallel to the background field of the LAPD on axis. The field geometry is shown in Fig.~\ref{fig:setup}b-c. In this configuration, far away from the dipole, its contribution goes to zero, leaving only the background field $B_z = B_0$ = 300G. Moving along the y-axis towards the dipole, $B_z$ decreases until it reaches a magnetic null-point at $y \approx -11$ cm, and then the negative contribution of the dipole dominates the field.\par
During the experiment, the magnetic fields are measured using a 3-axis, ten turn, 3~mm magnetic flux (bdot) probe \citep{everson09}, and the plasma potential is measured with an emissive probe \citep{martin15}. The probes are mounted on a 3D motorized drive and map out fields along an x,y,z grid in a 6~cm~$\times$~8~cm~$\times$~6~cm volume around the magnetic null point, with a distance of 2.5~mm between each position in the highest resolution runs. Measurements are repeated and averaged over three shots for every position of the probe, leading to volumetric datasets consisting of more than 36000 shots for both bdot and emissive probe runs. The typical laser shot-to-shot energy and pointing variations are below 5\% \citep{schaeffer18}.  A fast-gate ($\sim10$ ns) CCD camera captures time-resolved images of the experiment \citep{heuer17}. Swept Langmuir probes measured the initial ambient plasma density and temperature before the start of the laser-driven plasma.  \par

 
 \section{Experimental results}\label{sec:res}
The laser-produced plasma generates a fast-flow of carbon ions moving towards the target at a speed of 150~km/s (see Fig.~\ref{fig:fig2}a). Through collisionless coupling, the LPP transfers kinetic energy to the background hydrogen plasma \citep{bondarenko2017_a,bondarenko17_b}, which is set in motion and pushed towards the dipole.  The streak plot in Fig.~\ref{fig:fig2}a shows that the LPP expansion in the magnetized background also generates a magnetic compression front progressing at its velocity, while the magnetic field inside the plasma plume is expelled, creating a diamagnetic cavity \citep{cruz23,schaeffer18}. The strong currents supporting the expulsion of magnetic field inside the diamagnetic cavity can be seen at the bottom of Fig.~\ref{fig:fig2}c. As demonstrated in a previous experiment \citep{schaeffer22}, when the ram pressure of the incoming plasma is compensated by the dipole magnetic pressure, the LPP cannot progress forward and the magnetic compression associated with it stagnates and is reflected at $y=\simeq -9$~cm (see Fig.~\ref{fig:fig2}a), thus leading to the formation of an ion-scale magnetosphere ($D  = L_M/d_i \simeq 1.25 $, where $L_M$ is the standoff distance of the magnetosphere to the center of the dipole).\par
Figure~\ref{fig:fig2}b shows the total magnetic field $B_z$ as well as the magnetic field lines at $t = 0.65\ \mu$s in the Y-Z plane, which can be identified as the reconnection plane. Because of the plasma flow and magnetic compression associated with the LPP, the magnetic null point has been pushed forward by around 0.5~cm compared to its initial position, thus driving magnetic reconnection. A reconnection current sheet around the x-point can indeed be observed in Fig.~\ref{fig:fig2}e, which shows the current density $J_x$ in the Y-Z plane. The width of this current sheet is around 1.3~cm full-width half-maximum (FWHM), which corresponds to an intermediate scale between the ion and electron inertial lengths ($d_i = 7.2$~cm, $d_e = 0.2$~cm). We note that the current is estimated using magnetic data collected over many shots, so shot-to-shot fluctuations can contribute to widening the measured current sheet. Additionally, two electron outflows associated with this reconnection develop on both sides of the x-point, as can be seen through the current density $J_z$ shown in panel Fig.~\ref{fig:fig2}f.\par
Figure~\ref{fig:fig2}d shows the out-of-plane dynamic magnetic field $\Delta B_x$ in the reconnection plane. A quadrupolar shape of this magnetic field is observed around the x-point, which is a typical signature of Hall effects in reconnection, associated with spatial scales smaller than, or of the order of, the ion inertial length. In that case, the ions decouple from the electrons, leading to differentiated flows and thus currents that are the source of this structure \citep{shay98,mozer02,ren05,matthaeus05,brown06,uzdensky06}. This is observed in combination with a strong dipolar electrostatic field $E_y$ (see panel Fig.~\ref{fig:fig2}g), which can clearly be identified as the $\mathbf{J} \times \mathbf{B}$ Hall electric field. \par  
Finally, the electromagnetic energy dissipated in the plasma can be estimated by calculating the quantity $\mathbf{J}\cdot\mathbf{E}$, as shown in Fig.~\ref{fig:fig2}e, with $E = -\nabla \Phi_p - \partial\mathbf{A}/\partial t$ the total electric field taking into account the induced electromagnetic field computed from the magnetic data.  The electromagnetic fields associated with the LPP render the analysis of this quantity more complex, but one can clearly identify the dissipation peaking around the reconnection point, highlighting the magnetic energy being transferred to the plasma. \par

\begin{figure}[ht!]
\plotone{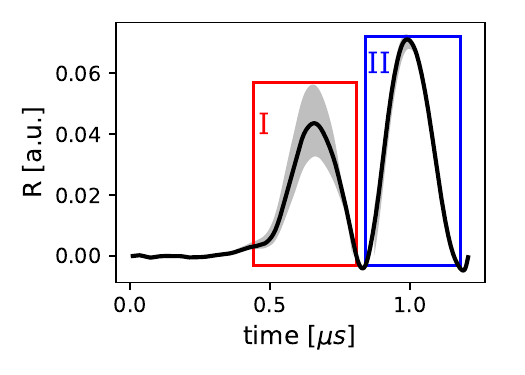}
\caption{ Normalized reconnection rate $R = E_{rec}/v_AB_0$. Zone I corresponds to actual reconnection, while zone II corresponds to expulsion of magnetic field by the LPP. The shaded area corresponds to an uncertainty on the position of the reconnection point of $\pm 2.5$~mm.
\label{fig:rec}}
\end{figure}

By taking advantage of the high-resolution volumetric data, enabled by the high repetition-rate of the experiment, we can calculate the reconnection rate by computing the annihilated magnetic flux through a Y-X surface with an edge passing by the reconnection point, similarly to \citet{greess21}. The reconnection electric field is then inferred using Faraday's Law: $-d\phi_B/dt = \oint \mathbf{E}\cdot\mathbf{dl}$. More details about the method to retrieve the reconnection rate are given in Appendix~\ref{sec:annexA}. Figure~\ref{fig:rec} shows the calculated normalized reconnection rate evolution in time. Two peaks are visible, identified as zones I and II. Only the first one corresponds to magnetic reconnection, while the second one is due to the expulsion of magnetic field by the laser-produced plasma. The reconnection rate peaks at $t = 0.67$ $\mu$s at a value of $R=0.04 \pm 0.01$ (corresponding to $E_{rec} = 2$~V/cm), which is of the order of the usual fast reconnection rate $R\sim0.1$ observed in collisionless systems \citep{yamada11,comisso16,cassak17}. This plot also indicates that, because of the transient and fast-driven nature of the experiment, reconnection does not reach a steady state, otherwise a saturation of R would be observed. Therefore, one could expect the reconnection rate to reach higher values in the case where it would be sustained for longer times.\par

\section{Particle-in-Cell Simulations} \label{sec:simulations}

\begin{figure*}[ht!]
\plotone{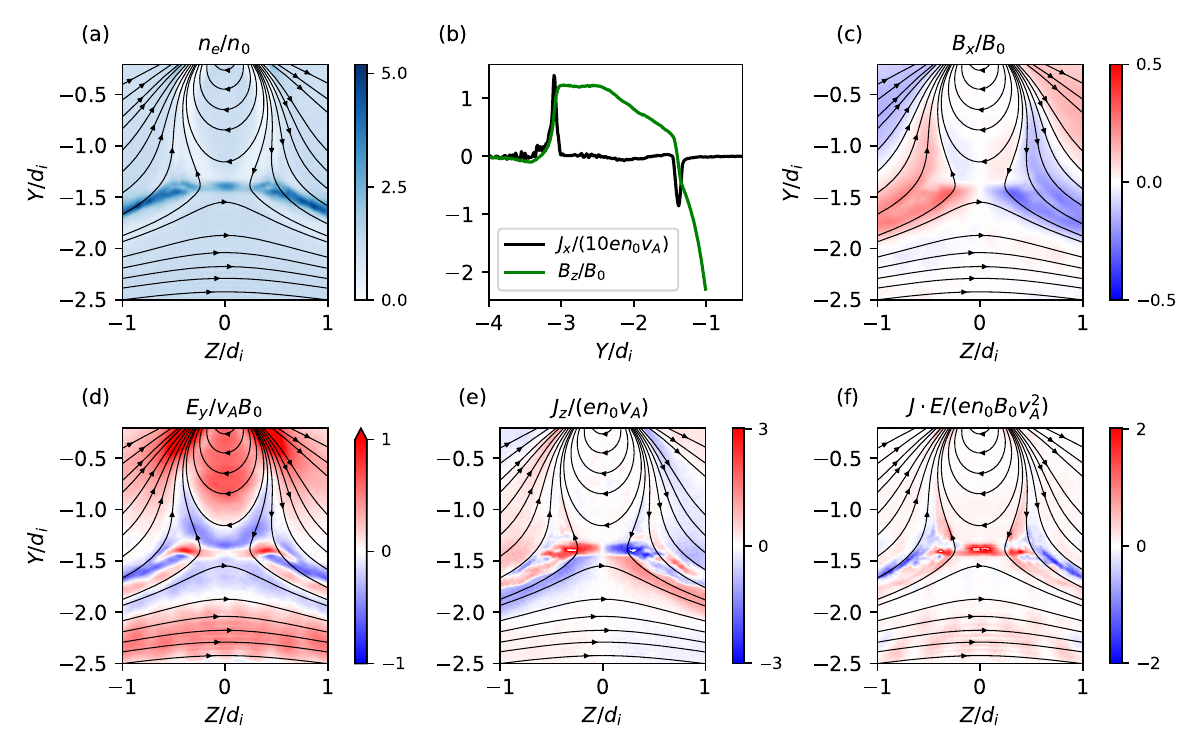}
\caption{Particle-in-cell simulation results at $\omega_{ci}t = 3.0$. (a) Electron density (blue) and magnetic field lines (black arrows). (b) Lineouts of magnetic field $B_z$ (green) and out-of-plane current density $J_x$ (black) along the y-axis at $z=0$. (c) Out-of-plane magnetic field $B_x$. (d) Electric field $E_y$ in the y direction. (e) Current density $J_z$ in the z direction. (f)  Dissipation of energy of the electromagnetic field into the plasma $\mathbf{J}\cdot\mathbf{E}$.  Quantities are normalized by a combination of the initial density $n_0$, initial magnetic field $B_0$, and/or Alfv\'{e}n speed $v_A$.
\label{fig:sim}}
\end{figure*}

In order to gain more insight about the micro-physics and kinetic effects that contribute to reconnection in the experiment, we carried out simulations using OSIRIS, an electromagnetic, fully-relativistic, massively parallel particle-in-cell (PIC) code \citep{osiris02,osiris13}. In the simulations, a carbon plasma slab modeling the laser-produced plasma moves towards a dipolar field through a region of magnetized ambient hydrogen plasma. The plasma and field parameters are set to be similar to the ones in the experiments, the speed of the plasma slab $v_{d0}$ is set so that the Alfv\'{e}nic Mach number is $M_A = 0.5$, and the dipole field strength is set so that the standoff distance is of the order of $L_M = 1~d_i$. In order to reduce numerical cost, reduced ion-electron mass ratios of $m_i/m_e =1200$ for the driver carbon plasma, and $m_i/m_e =100$ for the background hydrogen plasma are used, as well as a higher velocity ratio $v_{d0}/c = 0.1$. More details about the simulation setup are given in Appendix \ref{sec:appendixB}. \par
As the carbon slab progresses towards the dipole, its momentum is transferred to the initially at-rest background hydrogen plasma through collisionless coupling \citep{cruz23}. The magnetized ambient plasma is thus accelerated and pushed towards the dipole field. This inflow of plasma towards the magnetic null-point triggers reconnection of the fields lines. \par
Figure~\ref{fig:sim}a shows the magnetic field lines in the reconnection plane at a time where reconnection is occurring. A strong reconnection negative current in the out-of-plane (x) direction can be observed at the x-point at $y = -1.35~d_i$ in panel Fig.~\ref{fig:sim}b. This current structure width is of the order of the electron inertial length, highlighting the electron-only nature of reconnection happening here. The other current peak observed at $y=-3~d_i$ is the diamagnetic current from the driver slab that supports the expulsion of the magnetic field. These simulations also show the generation of significant Hall fields (see Fig.~\ref{fig:sim}c-d), similar to what is observed in the experiment, with quadrupolar $B_x$ and dipolar $E_y$ fields. Indeed, on a scale of the order of $d_i$, the electrons are frozen into the magnetic field lines but the ions are not. This leads to disparities in the flows of the two species that are source of currents that induce this quadrupole shape of the out-of-plane magnetic field. Associated with this reconnection, two anti-symmetric $J_z$ current structures develop on both the left and right sides of the x-point, corresponding to electrons out-flowing from the reconnection point at a super-Alfv\'{e}nic velocity $v_{out} = 2~v_A$.  Lastly, Fig.~\ref{fig:sim}f shows that the dissipation of electromagnetic energy in the plasma associated with reconnection is very localized to an electron-scale region around the reconnection point. \par
These simulation results demonstrate a physical behavior generally comparable to what has been observed experimentally, with a significant reconnection current driven by the fast flow associated with the laser plasma. The Hall fields observed experimentally and associated with differentiated flows between the ions and electrons are also well captured by the simulations. Still, it should be noted that reconnection features observed in the simulations tend to be thinner than the those measured on the LAPD, and typically match the electron inertial length. This is most probably due to a widening of the measured current sheet from limited spatial resolution and shot-to-shot variability in the experiment. \par

\section{Discussion} 
\label{sec:disc}
\subsection{Ohm's law analysis}
\label{sec:ohm}

In order to determine what physical phenomena is driving reconnection in the experiment, we evaluate the terms of the generalized Ohm's Law and compare their relative contribution to the reconnection electric field. The generalized Ohm's law can be written as follows \citep{hesse11}:

\begin{equation}
\mathbf{E} = - \mathbf{u}\times \mathbf{B}  +  \frac{\mathbf{J}\times \mathbf{B}}{n_ee} - \frac{\nabla p_e}{n_ee} - \frac{\nabla \cdot \mathbf{\Pi_e}}{n_ee}  + \frac{m_e}{e^2}\frac{d(\mathbf{J}/n_e)}{d{t}} 
\label{eq:ohm}
\end{equation}
\noindent where $\mathbf{E}$ is the electric field, and on the right-hand side: the first term is the contribution of the ion flow, the second term is the Hall electric field, the third is the term associated with the scalar electron pressure, and the fourth term is the non-diagonal pressure tensor term that accounts for pressure anisotropy ($\mathbf{\Pi_e} = \mathbf{P_e} - p_e\mathbf{I_3}$). Finally, the last term is the electron inertia term, where the total time derivative can be decomposed as: $\frac{d}{dt} = \frac{\partial}{\partial t}-(\mathbf{J}/(n_e e)) \cdot \nabla$.\par

\begin{figure}[ht!]
\plotone{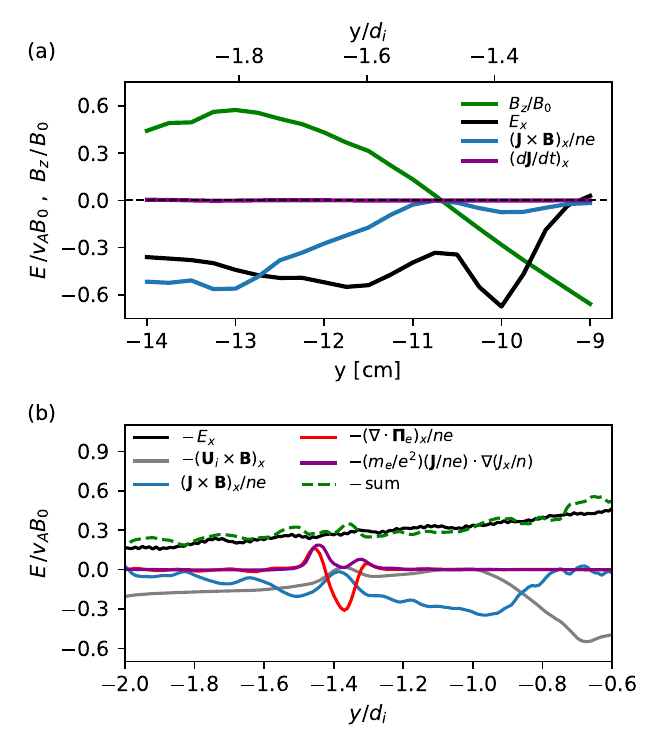}
\caption{Components in the x direction of the generalized Ohm's Law terms.  (a) In the experiment at $t = 0.65$~$\mu$s. The only accessible terms are the total electric field (black), the Hall term (blue), and the electron inertia term (magenta). The density is considered constant at $n_e = 1\times10^{13}$ cm$^{-3}$. All terms are normalized to $v_AB_0$. The normalized magnetic field $B_z/B_0$ (green) is shown as a reference. (b) In the simulation at $\omega_{ci}t = 3.0$. The green dashed line corresponds to the sum of the RHS terms of the generalized Ohm's law. The quantities are averaged over a zone of $1.5d_e$ in z, and smoothed along y by taking a moving average over $0.5d_e$. All terms are normalized to $v_AB_0$, and $E_x$ and the sum of the RHS terms are plotted with an opposite sign for clarity.
\label{fig:ohm}}
\end{figure}

The experimental data enable us to determine an estimate of the Hall term and the electron inertia term assuming a constant plasma density, as showed in Fig.~\ref{fig:ohm}a. It shows a significant contribution of the Hall term to the electric field $E_x$ on a typical scale of $d_i$ around the reconnection point, but it goes to zero at the magnetic null point. Therefore, other terms must be dominating the reconnection electric field on the electron scale. The PIC simulations give us access to kinetic information on the ions and electrons, and therefore gives us access to all the terms of the generalized Ohm's law, which are shown in Fig.~\ref{fig:ohm}b. We note that, since it is a 2D simulation, the out-of-plane electric field $E_x$ is purely electromagnetic, and the scalar electron pressure in the x direction is zero: $(\nabla p_e)_x =0$. This is not an issue because the scalar pressure primarily contributes to the electrostatic component of $\mathbf{E}$ and does not significantly contribute to the reconnection electric field \citep{greess21}. Similar to the experimental data, the simulations show that the Hall term is the main contribution to the non-ideal electric field on the typical $d_i$ scale, which corresponds to the ions breaking the frozen-in condition, as previously discussed. But in a electron scale zone around the reconnection point, the Hall term disappears and the reconnection electric field is driven solely by the electron pressure anisotropy. This pressure anisotropy is associated with the breaking of the frozen-in condition by the electrons as the magnetic field vanishes on the x-point, and the field reversal on each side leads to an electron-meandering motion in the y direction around the magnetic null point that leads to a non-gyrotropic velocity distribution \citep{hesse99,ishizawa04,ishizawa05,ng11}. The meandering width $l_{me}$ corresponds to the distance at which the local Larmor radius of the electrons is equal to the distance to the x-point, $\rho_e(y) = y$, and can be approximated using the magnetic field gradient scale $L_B = B_0/B'_y$ as $l_{me} = (\rho_e L_B)^{1/2}$, where $\rho_e = m_ev_{the}/(eB_0)$ \citep{nakamura16}. In the simulations, the value of the meandering scale is around $l_{me} = 0.4~d_e$ and approximately matches the half-width of the non-gyrotropic electric field. It appears that the current sheet width is mainly dictated by the pressure anisotropy and therefore by this electron meandering scale. In the experiment, because the magnetic field gradient scale is larger, this meandering scale is greater than the electron inertial length $l_{me,exp} \sim 1.8~d_e \sim 3.2$~mm, which could also explain why the current sheet is wider than in the simulations. Additionally, Fig.~\ref{fig:ohm}b shows a finite electron inertial term on the scale of the electron inertial length around the reconnection point; however, simulations with a higher mass ratio ($m_i/m_e = 900$) indicate that the importance of this term decreases, and we expect this term to become negligible when using a fully realistic mass ratio.\par

\subsection{Application to lunar reconnection}
\label{sec:moon}

Our laboratory experiment operates in a physical regime similar to lunar mini-magnetospheres associated with local crustal magnetic anomalies, and thus can provide insights into the physics driving lunar reconnection events such as the one identified in the recent work of \citet{sawyer23}. Table~\ref{tab:param} shows a comparison of important dimensionless quantities between the experiment, PIC simulations, and lunar mini-magnetospheres. Notably, our laboratory experiment reproduces the collisionless, ion-scale, low-$\beta$ regime that can be found on the Moon. Therefore, it is probable that the effect of Hall physics associated with the different electron-ion scales would also be observed in lunar mini-magnetospheres. Indeed, \citet{sawyer23} observed Hall $\mathbf{J}\times \mathbf{B}$ electric fields in satellite measurements in regions of lunar mini-magnetosphere associated with magnetic reconnection. Additionally, the combination of high Lundquist number and small system size places both systems in the collisionless single X-line reconnection regime identified by \citet{ji11}. \par

\begin{table}[ht!]
\vspace{0.5cm}
\centering 
\begin{tabular}{|c | c c c| } 
\hline 
Param. & Lunar mini-mag & LAPD exp. & PIC sim. \\ [0.5ex] 
\hline 
$L_M/d_i$ & 0.3 - 5 (0.3 in [\citeyear{sawyer23}]) & 1.25 & 1 \\ 
$L_M/\rho_i$ & 0.1-1 (0.1 in [\citeyear{sawyer23}]) & 1.7 & 2 \\
$M_A$ & 4-8 & 0.75 & 0.5 \\
$L_M/\lambda_{ii}$ & $\ll1$ & 0.02 & - \\
$L_M/\lambda_{ee}$  & $\ll1$ & 2 & - \\ 
S & $\gg1 \ (10^8)$ & 550 & - \\ 
$\beta$  & 0.4 & 0.02 & 0.002 \\  [1ex]
\hline 
\end{tabular}
\caption{Comparison of dimensionless parameters for lunar and LAPD mini-magnetospheres, as well as collisionless PIC simulations. $d_i$ is the ion inertial length, $\rho_i$ the ion gyroradius, $M_A$ the Alfv\'{e}nic Mach number, $\lambda_{ii}$ and $\lambda_{ee}$ are the ion-ion and electron-electron collisional mean-free-path, respectively, $S$ is the Lundquist number, and $\beta$ is the plasma beta.} 
\label{tab:param} 
\end{table}

Nevertheless, two significant differences between the laboratory and lunar magnetospheres worth addressing can be identified. In the lunar case, the ion gyroradius is much larger than the system size, rendering them completely unmagnetized.  In the experiment, while the ions are still only weakly magnetized, the ion gyroradius is slightly smaller than the system size, so the impact of the magnetic field on them could be amplified compared to the lunar case. The other discrepancy is that we drive a sub-Alfv\'{e}nic flow in the laboratory, while the solar wind impinging on the moon is mostly super-Alfv\'{e}nic. This notably prevents the possible formation of a shock, and leads to a generally weaker interaction. Still, it can be noted that the Moon's orbit around the Earth also leads it to cross Earth's magnetosphere, where it would experience sub-Alfv\'{e}nic conditions \citep{liuzzo21}. \par

\section{Conclusions}

We have presented experimental results on magnetic reconnection in mini-magnetospheres, where we identify electron-only magnetic reconnection through the observation of a sub-ion scale reconnection current structure, as well as electron outflows on each side of the x-point. Additionally, the measured electric and magnetic field exhibit markers of Hall physics at play in ion-scale magnetospheres. The normalized reconnection rate has been estimated to be around $0.04\pm 0.01$.\par

Particle-in-cell simulations successfully reproducing the experimental features were carried out and enabled us to study the underlying mechanisms driving reconnection in the experiment, through the evaluation of the generalized Ohm's law terms. In accordance with the experimental observations, the Hall term dominates the non-ideal electric field on an ion-scale around the reconnection point. But this Hall electric field vanishes in an electron-scale region around the x-point and its contribution is replaced by the electron pressure anisotropy term, which drives reconnection on this electron scale. \par  

By comparing dimensionless quantities relevant to magnetic reconnection in the experiment and in lunar mini-magnetospheres, we have established that our experimental and numerical study is relevant to understanding the physics driving magnetic reconnection on the Moon, providing important information on the nature of reconnection in a lunar setting where low-altitude measurements are scarce and incomplete.  These results provide experimental demonstration that magnetic reconnection can occur in ion-scale lunar magnetospheres and can validate recent observational \textit{in situ} evidence from the THEMIS-ARTEMIS mission of such a reconnection event on the Moon \citep{sawyer23}. Our data shows that important Hall electric and magnetic fields develop in association with magnetic reconnection, and that this reconnection is solely supported by the electrons. The numerical modeling of the experiment using PIC simulations indicates that the electron pressure anisotropy plays a fundamental role in driving this electron-scale reconnection. To explore this kinetic regime in more detail, we will conduct further experiments aiming at probing additional terms in the generalized Ohm's law. 

\begin{acknowledgments}
The authors are grateful to the staff of the BaPSF for their help in carrying out these experiments. The experiments were supported by the NSF/DOE Partnership in Basic Plasmas Science and Engineering Award Nos. PHY-2010248 and PHY-2320946, and by the Defense Threat Reduction Agency and Lawrence Livermore National Security LLC under Contract No. B655224, and the Naval Information Warfare Center-Pacific (NIWC) under contract NCRADA-NIWCPacific-19-354. The simulations were supported by FCT (Portugal) under the Project X-MASER No. 2022.02230.PTDC and the Grant APPLAuSE UI/BD/154620/2022. The simulations were performed at LUMI (Kajaani, Finland) within EuroHPC-JU Project No. EHPC-REG-2021R0038, and at the IST cluster (Lisbon, Portugal). 
\end{acknowledgments}

\appendix

\section{Reconnection Rate}
\label{sec:annexA}
\begin{figure}[ht!]
\centering
\includegraphics[width=0.3\textwidth]{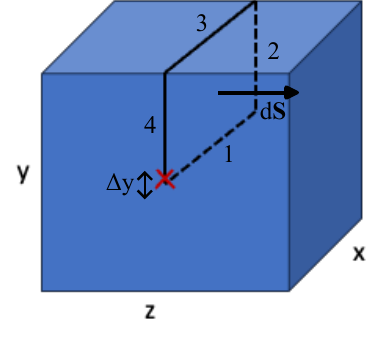}
\caption{Principle of the reconnection rate calculation. The blue box represents the whole data volume, the red cross the magnetic null point position in (y,z), and the rectangle is the area on which the magnetic flux is computed.
\label{fig:annexA}}
\end{figure}

The reconnection rate can be obtained by computing the annihilated magnetic flux through a Y-X surface with an edge passing by the reconnection point described in Fig.~\ref{fig:annexA}. This surface extends from both -x,+x ends of the data volume, and from the magnetic null point to the top edge of the data box in y. That way, we can calculate the variation (annihilation) of magnetic flux on the top side of the x-line : $-\frac{d\phi_B}{dt} = \iint \mathbf{B}\cdot d\mathbf{S}$. Using Faraday's Law, this magnetic flux is linked to the electromagnetic part of the electric field on the edges of the surface: $-\frac{d\phi_B}{dt} = \oint \mathbf{E}\cdot d\mathbf{l}$. \par

To recover the reconnection electric field $E_{rec} = E_{1}$ we make several assumptions: (i) we consider the line-averaged electric field, so that the integral becomes: $\oint \mathbf{E}\cdot d\mathbf{l} = (E_1 - E_3) \Delta x + (E_2-E_4) \Delta y$; (ii) the problem is symmetric along x, so $(E_2-E_4) = 0$; and (iii) the induced electric field along edge 3 is zero $E_3 = 0$, so Faraday's Law becomes  $-\frac{d\phi_B}{dt} = E_{rec}\cdot \Delta x$. The last assumption requires additional justification. Indeed, it is not verified at all times, but using the time-resolved data collected, we determine that there is no time-varying magnetic field on the top edge of the data box until $t\geq0.9~\mu$s, and therefore no induced electric field either. So this assumption is valid during the first peak of Fig.~\ref{fig:rec} at $t = 0.65~\mu$s, which corresponds to the actual reconnection. The uncertainty on the reconnection rate (shaded area on Fig.~\ref{fig:rec}) is obtained by varying the estimated position of the reconnection point by $\pm\Delta y=$ 2.5~mm, which is the measurement resolution. \par

\section{Simulations Setup}
\label{sec:appendixB}
\par The 2D PIC simulation presented in this work stems from a simplified description of the experimental setup at the $x=0$ plane. In the simulation, a unmagnetized driver plasma moves against a background plasma permeated by an internal uniform magnetic field $\mathbf{B_0}$ and an external dipolar magnetic field $\mathbf{B_{d}}$. The simulations consist of a $12~d_i \times 12~d_i$ region, with 1200$\times$1200 cells, and open and periodic boundaries in $y$ and $z$, respectively. The simulation considers 25 particles per cell for the driver's ions and electrons and 64 for the background's. The simulation spatial units are normalized to the background ion-skin depth $d_i$ and the temporal units to the background ion gyroperiod $1/\omega_{ci}$.

\par The driver plasma moves in the $y$-direction with a flow velocity of $v_{d0}=0.1\ c$ and it is located between $-6<y/d_i<-4$, while the background is located between $-4<y/d_i<4$. Both plasmas have an infinite width, and a vacuum region exits between $-8<y/d_i<-6$. To simulate the carbon target, the driver ions have a charge of +4 and an ion-to-electron mass ratio of $m_{id}/m_e=1200$, while the background ions have a charge of +1 and mass ratio of $m_{i0}/m_e=100$. The plasmas have uniform and equal ion density $n_{d0}={n_0}$ and an electron thermal velocity of $v_{thex}=v_{they}=v_{thez}=v_{d0}$. The ions are considered to be in thermal equilibrium~\citep{cruz22}. 

\par $\mathbf{B_0}$ and $\mathbf{B_{d}}$ are defined in-plane and are antiparallel in the $z$-direction. $\mathbf{B_0}=B_0\ \mathbf{\hat{z}}$ was chosen such that the Alfénic Mach number $M_A=v_{0d}\sqrt{\mu_0 n_0m_{i0}}/B_0$ is equal to 0.5. The dipole is centered at $\{y,z\}=\{0,0\}$ and the dipolar magnetic moment was chosen such that the magnetosphere's standoff distance is $L_M = d_i$, leading to an initial null magnetic field at $y\approx -1.5\ d_i$.




\end{document}